\documentclass[onecolumn,preprintnumbers,amsmath,amssymb]{revtex4}
\usepackage{graphicx}
\usepackage{dcolumn}
\usepackage{bm}
\usepackage{times}

\begin{document}
\title{Two Distinct Phases of Bilayer Graphene Films on Ru(0001)}

\author{Marco Papagno$^{1,2\ast}$, Daniela Pacil\'e$^{1, 2}$, Dinesh Topwal$^{3,4}$, Paolo Moras$^{1}$, Polina Makarovna Sheverdyaeva$^{1}$, Fabian Donat Natterer$^{5}$, Anne Lehnert$^{5}$, Stefano Rusponi$^{5}$, Quentin Dubout$^{5}$,
Fabian Calleja$^{5}$, Emmanouil Frantzeskakis$^{5}$, St\'ephane Pons$^{5,6}$, Jun Fujii$^{7}$, Ivana Vobornik$^{7}$, Marco Grioni$^{5}$, Carlo Carbone$^{1}$,  and Harald Brune$^{5}$}
\affiliation{
$^{1}$~\mbox{Istituto di Struttura della Materia, Consiglio Nazionale delle Ricerche, Trieste, Italy} \\
$^{2}$~\mbox{Dipartimento di Fisica, Universit$\acute{a}$ della Calabria, 87036 Arcavacata di Rende (Cs), Italy} \\
$^{3}$~\mbox{International Center for Theoretical Physics, Trieste, Italy}\\
$^{4}$ ~\mbox{Institute of Physics, Sachivalaya Marg, Bhubaneshwar, 751005, India}\\
$^{5}$~\mbox{Institute of Condensed Matter Physics (ICMP), Ecole Polytechnique F\'ed\'erale de Lausanne (EPFL) , CH-1015 Lausanne, Switzerland}\\
$^{6}$~\mbox{Institut des NanoSciences de Paris (INSP), Universit{\'e} Pierre et Marie Curie (UPMC), CNRS UMR 7588, France}
$^{7}$~\mbox{CNR-IOM, TASC Laboratory, Basovizza, Trieste, Italy}\\
}

\maketitle
{\bf ABSTRACT: By combining angle-resolved photoemission spectroscopy and scanning tunneling microscopy we reveal the structural and electronic properties of multilayer graphene on Ru(0001). We
prove that large ethylene exposure allows to synthesize two distinct phases of bilayer graphene with different properties. The first phase has Bernal AB stacking with respect to the first graphene layer, displays weak vertical interaction and electron doping. The long-range ordered moir{\'e} pattern modulates the crystal potential and induces replicas of the Dirac cone and minigaps. The second phase has AA stacking sequence with respect to the first layer, displays weak structural and electronic modulation and {\it p}-doping.
The linearly dispersing Dirac state reveals the nearly-freestanding character of this novel second layer phase.}

The peculiar electronic properties of graphene result from its two-dimensional (2D) honeycomb crystal lattice.~\cite{nov04, nov05, zha05, ber06, oht06, nov07} This 2D structure makes every carbon atom a surface atom and thus strongly exposed to the surrounding environment. The simple act of resting on a substrate may modify the electronic band structure and cause a precipitous decline of its charge carrier mobility.~\cite{oht06, oos07, fei09, rut11} For example, when graphene is placed on top of another graphene layer in a Bernal (ABAB) stacking the coupling between the planes modifies the linear dispersion of the massless charge carriers into a hyperbolic dispersion of massive chiral fermions.~\cite{gei07,mcc06}
The graphene layers can be partially decoupled by misorienting them by more than 20$^{\circ}$~\cite{lui11} or by an AA stacking sequence.~\cite{mak10, lob11} In these cases, the band structure has been predicted to be the superposition of two single-layer bands of freestanding graphene and only states with linear dispersion close to the Fermi level in proximity to the K points are present. 

\begin{figure}
\includegraphics[width = 9 cm]{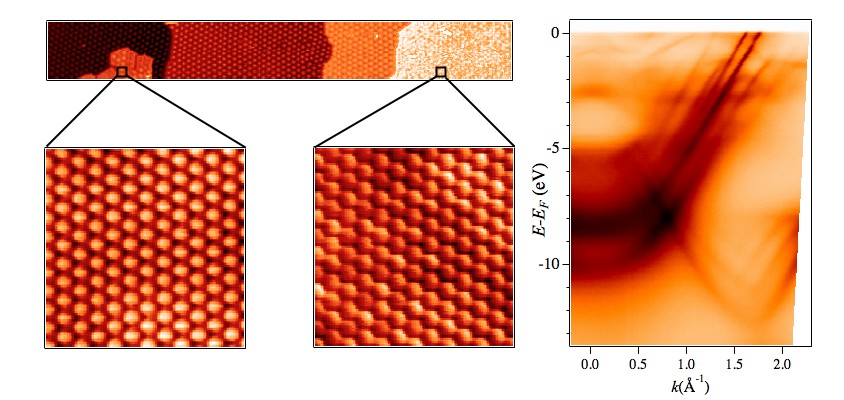}
\end{figure}

Yet another way to preserve the electronic properties of graphene is to adsorb it on a material that has only very weak interaction with graphene. Recently, graphene supported on a hexagonal boron nitride substrate~\cite{dea10} showed electrical properties comparable to that of freestanding graphene. 
On SiC(0001)~\cite{oue10, bos07, haa08} and on Ni~\cite{var08, yu08, rei09} the first graphene layer saturates the substrate bonds and allows the synthesis of subsequent graphene layers with nearly-freestanding character. 

Another surface that strongly modifies graphene properties is Ru(0001). In this case, the first graphene layer relieves the strain due to the lattice mismatch between graphene and the substrate by developing a $(23 \times 23)$ moir{\'e} structure.~\cite{mar08} Core level spectroscopy,~\cite{pre08} scanning tunneling microscopy  (STM)~\cite{mar07, mar08,wan08, mor10, str11, fen11} experiments and density functional theory calculations~\cite{wan08, bru09, jia09, str11, fen11, ian11} have revealed spatial inhomogeneities in the charge density reflecting regions with varying graphene-substrate hybridization. Angle-resolved photoemission spectroscopy (ARPES) studies displayed a single and broad $\pi$ band with a relatively large gap.~\cite{bru09}  
The first graphene layer saturates the metal bonds and acts as a template for the growth of a subsequent graphene layer with freestanding character~\cite{sut08, esu09} which displays a sharp and linearly dispersing Dirac cone.~\cite{sut09}

\begin{figure*}
\includegraphics[width = 19 cm]{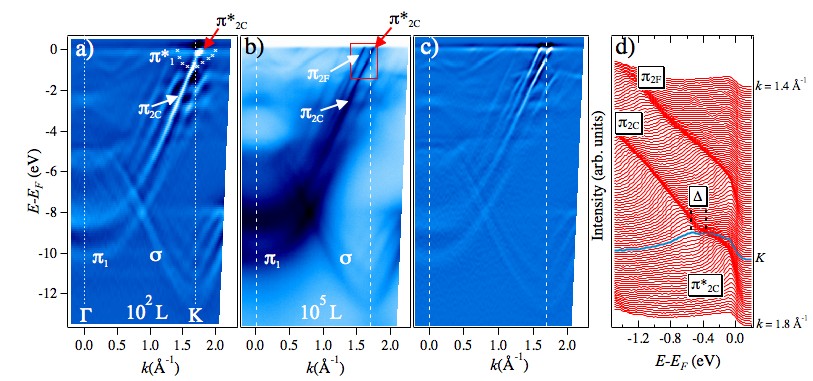}
\caption{(Color online) (a) Second derivative of the ARPES intensity map along the $\Gamma$${\rm K}$ direction of graphene for the low-exposure sample.  $\pi_{\rm 1}$ and $\pi_{\rm 2C}$ label the electronic states of the first monolayer and of the corrugated second monolayer graphene, respectively. The broad feature enclosed by white crosses is ascribed to the $\pi_{\rm 1}^{\ast}$ state strongly modified by the substrate interaction. Due to marginal electron doping of the corrugated layer we observe also the $\pi_{\rm 2C}^{\ast}$ state. (b) Raw and (c) second derivative ARPES intensity maps for the high-exposure sample.
 $\pi_{\rm 2F}$ identifies a new electronic state of the second monolayer. (d) Energy distribution curves from the spectra enclosed by the red rectangle in panel (b). $\Delta$ labels the electronic gap between the $\pi_{\rm 2C}$ and $\pi_{\rm 2C}^{\ast}$ states.
}
\label{arpes}
\end{figure*}

Here, we provide evidence of 
 a novel second layer phase on Ru(0001), which arises after exposure to more than 10$^{5}$~L ethylene that coexists with the hitherto reported phase. 
We compare ARPES data of the two second layer phases and find that the new  phase  gives rise to a second linearly dispersing $\pi$ band which is shifted with respect to the one of the first phase due to significant {\it p}-doping. Our STM data reveal that the two second layer phases have different stacking sequence and moir{\'e} amplitude.

{\bf RESULTS AND DISCUSSION}

Figure~1(a) displays the second derivative of the ARPES intensity map along the $\Gamma$${\rm K}$ direction of the graphene Brillouin zone after exposure to 10$^{2}$~L ethylene. We reproduce the previously observed relatively broad $\pi_{\rm 1}$ and $\pi_{\rm 1}^{\ast}$ bands and the sharp $\pi_{\rm 2C}$ and  $\pi_{\rm 2C}^{\ast}$ states. In addition, we  identify replica bands associated to the $\pi_{\rm 2C}$ and $\sigma$ states. 
In accordance with literature,~\cite{sut09} we assign the bands as follows. 
The monolayer interacts strongly with the underlying substrate and displays a parabolic like dispersion of the $\pi$ state~\cite{bru09, sut09, end10} ($\pi_{1}$ band). This band has the minimum at $E-E_{F}$=-10.1~eV at the $\Gamma$ point and the maximum at $E-E_{F}\simeq$-4~eV at the {\rm K} point. 
The strong interaction with the substrate also modifies the $\pi_{\rm 1}^{\ast}$ into a diffuse and faint band (enclosed by the white crosses in Figure~1(a)) in proximity to the Fermi level. The first layer saturates the metal orbitals and allows the growth of a second graphene layer which exhibits a sharp and {\it n}-doped $\pi_{\rm 2C}$ Dirac cone, with the Dirac point at 0.45$\pm$0.05~eV below {\rm E$_{F}$}, along with replica and $\sigma$ bands.~\cite{sut08, esu09, sut09, end10}  As will become clear from the STM data presented below, we refer to this layer as corrugated layer, therefore the label C.

\begin{figure*}
\includegraphics[width = 14 cm]{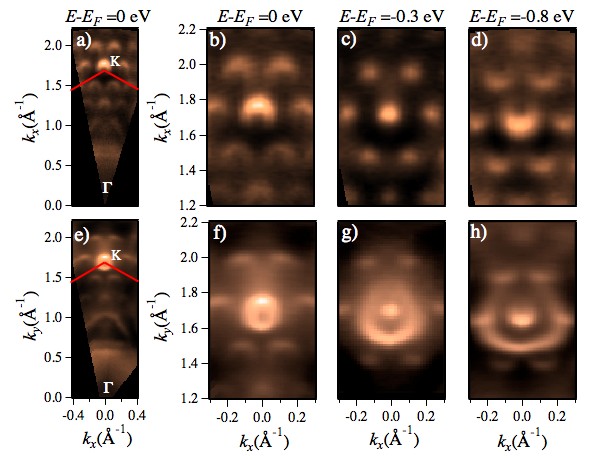}
\caption{(Color online) 
(a) Constant energy maps  of graphene states and (b) close-up of the  {\it K} point for the 10$^{2}$~L sample at $E-E_{F}$=0. Panels (c) and (d) display underlying bands for  $E-E_{F}$=-0.3~eV and -0.8~eV, respectively. (e) Fermi surface of graphene bands (f) and zoom-in around the  {\it K} point for the 10$^{5}$~L sample. Panels (g) and (h) show the constant energy maps measured at $E-E_{F}$=-0.3~eV and -0.8~eV. Red lines in (a) and (e) mark the limits of the  first Brillouin zone of graphene.
}
\label{arpes}
\end{figure*}

\begin{figure}
\includegraphics[width = 8.6 cm]{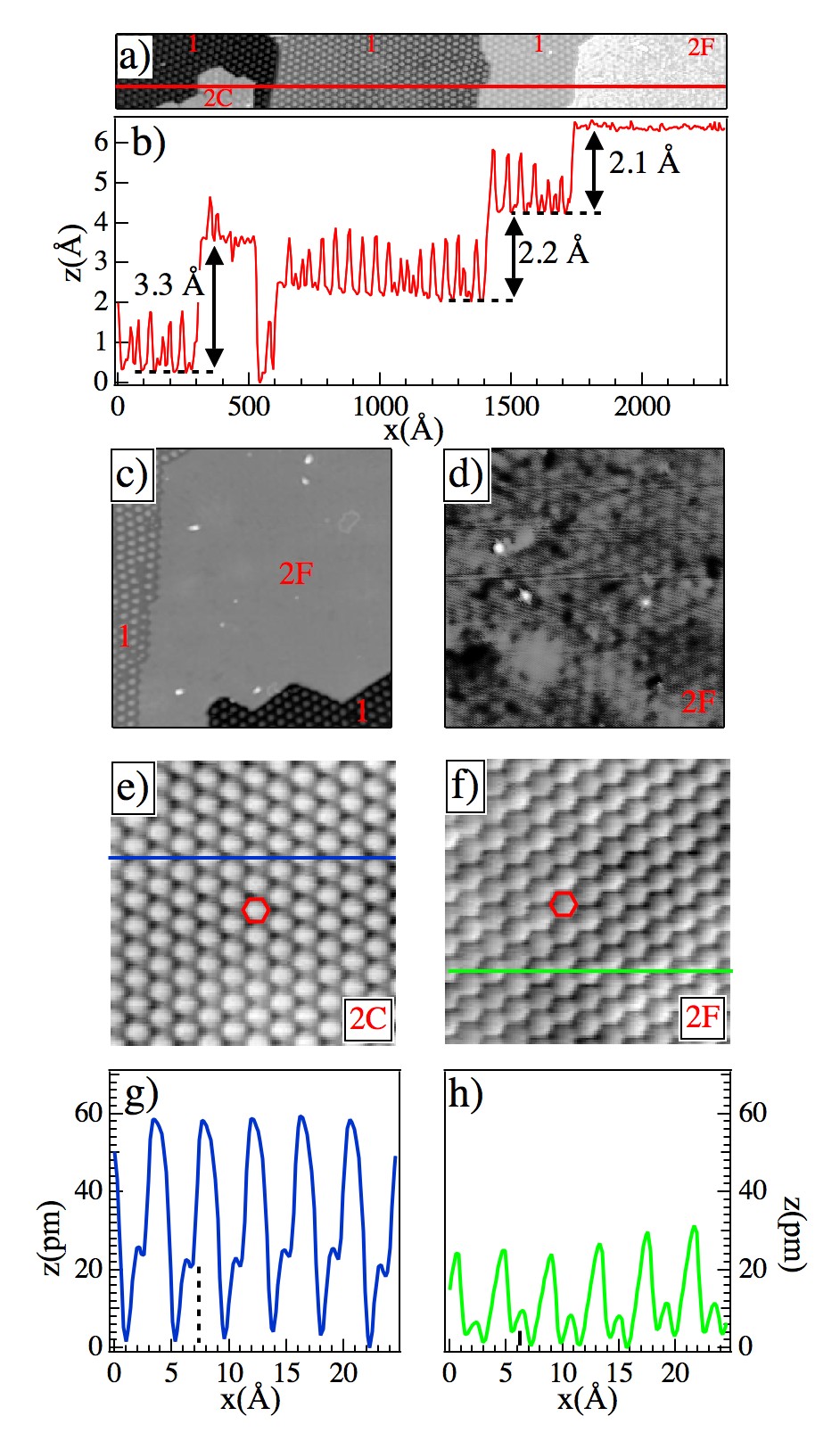}
\caption{(Color online) (a) Constant current STM image of graphene on Ru(0001) (ethylene exposure 10$^{5}$~L, {\rm V$_{t}$}=-3~{\it V}; {\rm I$_{t}$}=75~{\it pA};  T=4~K; size 2320~\AA$\times$283~\AA). 1, 2C, and 2F, stand for first layer, and second corrugated and flat layer, respectively. (b) Apparent height profile along the red line displayed in (a). (c)~Constant current STM image of a sample region with coexisting first graphene layer and 2F layer ({\rm V$_{t}$}=-3~{\it V}; {\rm I$_{t}$}=75~{\it pA}; T=4~K; size 762~\AA$\times$762~\AA). (d) STM image of the 2F island ({\rm V$_{t}$}=-1~{\it V}; {\rm I$_{t}$}=100~{\it pA}; T=4~K; size = 391~\AA$\times$391~\AA) showing a very weak moir{\'e} pattern.
(e) High-resolution STM image for 2C ({\rm V$_{t}$}=-0.1~{\it V}; {\rm I$_{t}$}=200~{\it pA}; T=4~K; size 25~\AA~$\times$ 25~\AA) and (f) 2F ({\rm V$_{t}$}=-0.~{\it V}; {\rm I$_{t}$}=100~{\it pA}; T=4~K; size 25~\AA~$\times$ 25~\AA). Note that the STM tip gives with respect to the atomic geometry a reversed contrast with C atoms imaged as dark spots. Red hexagons mark the graphene honeycomb lattice. (g) and (h) Apparent height profiles along the blue and green lines displayed in (e) and (f), respectively. Dotted and continuous vertical bars in (g) and (h) highlight the difference in the apparent height between carbon atoms with different sublattice symmetry for the 2C and 2F layers, respectively.
}
\label{STM}
\end{figure}

\begin{figure}
\includegraphics[width = 9 cm]{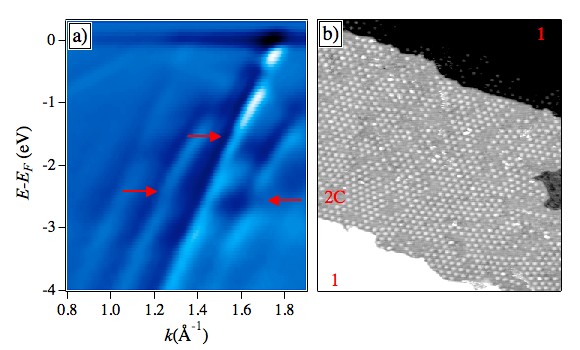}
\caption{(Color online) (a) First derivative of the ARPES intensity map measured 2$^{\circ}$ off the $\Gamma${\rm K} direction of the 10$^{2}$~L sample. The arrows highlight break in the bands dispersion revealing minigaps at the band crossing. 
(b) The STM image on the same sample shows a moir{\'e}  pattern with marginal long-range order of the corrugated layer  ({\rm V$_{t}$}=0.88~{\it V}; {\rm I$_{t}$}=170~{\it pA}; T=300~K; size 950~\AA~$\times$~1060~\AA).}
\label{arpes}
\end{figure}

Figures~1(b) and 1(c) show raw and second derivative ARPES maps recorded  after exposing the 10$^{2}$~L surface to further 10$^{5}$~L ethylene. In addition to the first and corrugated second monolayer  bands, we find a new $\pi$ state, which we call $\pi_{\rm 2F}$. F stands for flat layer, see below. The main features of this state are better captured in Figure~1(d) which displays energy distribution curves (EDCs) of the region marked by  the red rectangle in Figure~1(b). 
The $\pi_{\rm 2F}$ state disperses upward parallel to the $\pi_{\rm 2C}$ state and crosses the Fermi level before the maximum is reached reminiscent of  {\it p}-doping. By extrapolation of its linear dispersion we estimate the maximum of this band at 0.60$\pm$0.05~eV above the Fermi level. The constant $dE/dk$ gradient confirms the nearly-free character of the layers and yields an electron group velocity of $v_g = (1.0\pm0.1) \times 10^{6}$~m/s for both bands. 
Furthermore, EDC spectra highlight an electronic gap $\Delta$ between the $\pi_{\rm 2C}$ and $\pi_{\rm 2C}^{\ast}$ states of 200$\pm$30~meV at the {\rm K} point (blue curve). This gap is also observed in the 10$^{2}$~L sample at variance with earlier measurements.~\cite{sut09, end10}

Figure~2 presents constant energy maps close to the {\it K} point of the graphene Brillouin zone and acquired on the low exposure  (Figures~2(a)-2(d)) and high exposure samples (Figures~2(e)-2(h)). 
The Fermi surface of the 10$^{2}$~L sample displays the Dirac-like electron pocket of the  $\pi_{\rm 2C}$ state centered at the {\it K} point,  and surrounded by hexagonally arranged replicas (Figures~2(a)-2(b)). The excellent long-range order of the surface allows to observe up to the third  replica and to accurately measure the spacing between the replicas, amounting to $0.24 \pm 0.02$~\AA$^{-1}$. 
The electron density estimated from the relative size of the graphene Fermi surface with respect to the surface of Brillouin zone of graphite is 0.9$\times10^{13}$ cm$^{-2}$. For $E-E_{F}$=-0.3~eV (Figure~2(c)), graphene bands are point-like, and then display hole pockets for  $E-E_{F}$$<-$0.6~eV (Figure~2(d)). 
Similarly to graphene on SiC(0001),~\cite{oht06} these replicas could be induced by diffraction effects due to
the moir{\'e} of the first graphene layer,~\cite{sut09} or as discussed later in the text, to the superlattice potential of the second graphene layer.    

The Fermi surface of the 10$^{5}$~L sample in Figures~2(e) and 2(f) shows an additional Dirac-like hole pocket due to the  $\pi_{\rm 2F}$ band coexisting with the bands of the corrugated layer. The charge carrier concentration estimated from the relative size of the  $\pi_{\rm 2F}$ hole pocket is 2.5$\times10^{13}$ holes per cm$^{-2}$. With decreasing energies (Figures~2(g)-2(h)) the $\pi_{\rm 2F}$ state spreads out in {\it k}-space without affecting the $\pi_{\rm 2C}$ state.

It is evident from comparison of the constant energy maps of both samples that the features of the 10$^{2}$~L sample appear at the same energies and positions in {\it k}-space also on the 10$^{5}$~L sample. Therefore the additional $\pi_{\rm 2F}$ feature exhibited by the 10$^{5}$~L sample coexists but does not interact with the $\pi_{\rm 2C}$ features. This is a notable difference with respect to multilayer graphene where the number of $\pi$ bands increases with the number of layers, but their dispersion gets modified by interlayer coupling.~\cite{oht06, sut09} The absence of this interlayer coupling in the present case strongly suggests the coexistence of two second layer phases rather than the coexistence of second and a third monolayer. This conjecture is unequivocally proven by the following STM results recorded on a sample for which identical ARPES results as the ones shown in Figures~2(e)-(h) have been obtained. 
 Figure~3(a) shows an STM image of the 10$^{5}$~L sample. The Ru surface is entirely covered by a well-ordered graphene first layer. This layer displays the  $(23 \times 23)$ moir{\'e} pattern~\cite{mar08, pre08, bru09, sut09, sut08} resulting from the lattice mismatch between Ru and graphene.~\cite{mar08,vaz08,mor10} Part of the first layer is covered by second layer graphene. On large scale images we clearly discern two types of second layer  by their different moir{\'e} corrugation. A patch of the second layer type already reported in literature~\cite{esu09} is visible in the lower left part of the image. It displays a moir{\'e}  pattern with a maximum corrugation of 1.10~$\pm$0.05~\AA. It is therefore referred to as corrugated (C) layer  and covers 12$\pm$2~\% of the surface. The second type of the second layer is visible on the right hand side of Figure~3(a). It covers the 25~$\pm$5~\% of the first layer. Panel~(c) shows that the 2F layer is essentially flat compared to the first monolayer. The STM image in Figure~3(d) shows the weak moir{\'e} contrast of the 2F layer. The apparent height amplitudes of the moir{\'e} patterns of the first, and 2F layers are 1.30~$\pm$0.05, and 0.05~$\pm$0.02~\AA, respectively.

The line profile in Figure~3(b) demonstrates that 2C and 2F further distinguish themselves by their apparent heights with respect to the monolayer of 3.3 $\pm$ 0.1~\AA~and 2.1 $\pm$ 0.1~\AA, respectively. The apparent height of a Ru(0001) step is 2.2 $\pm$ 0.1~\AA, in agreement with the literature value of 2.14~\AA. The value measured for 2C corresponds to the distance of 3.3~\AA~between the atomic 
 layers in graphite which is Bernal, {\it i.e.}, AB stacked.~\cite{esu09, sun10} The AB stacking breaks the graphene sublattice symmetry while the AA stacking does not. Therefore high-resolution STM data can discern the two stackings, as evidenced by Figures~3(e) and (f) with their line profiles. 
In Figure~3(e) the protrusions mark the centers of the C$_6$ rings and the two carbon atoms per graphene unit cell are imaged as dark spots; The line profile taken along the blue line and displayed in Figure~3(g) evidences a difference in apparent height of 25 $\pm$ 6~{\it pm}  between these two atoms (vertical dotted line in Figure~3(g)). This lifting of the degeneracy between the A and B sublattices takes place in all the investigated regions of the corrugated layer and is responsible for the bandgap $\Delta$ between the $\pi_{\rm 2C}$ and $\pi_{\rm 2C}^{\ast}$ states observed in Figure~1. 
\newline The apparent height  between the first and the 2F layer is considerably smaller than the interlayer distance in graphite. However, in a previous STM work an apparent height of 2.5~\AA~has been observed between the first  and the second graphene layer on SiC.~\cite{rut07} High resolution transmission electron microscope studies reported a distance of 2.2~\AA~between two graphene layers grown on the (111) surface of diamond.~\cite{lee08} This relatively small interlayer distance has been explained  by an AA$^{\prime}$ stacking, where alternate planes are translated by half the hexagonal cell width. In our case  a shift between adjacent layers can be excluded since a detailed analysis of the atomic positions of first and flat second layer shows that they are in registry.~\cite{cal}

The AA stacking between the first and the 2F layer is further supported by the within the error bar identical apparent height of the two graphene sublattices reported in Figure~3(f). C atoms are again imaged  dark. The apparent height difference between the two C sublattices in 2F layer has a mean value of 4$\pm$7~{\it pm} (vertical bar in Figure~3(h)). Therefore the STM data reveal a sublattice degeneracy for the 2F identical to free standing graphene while this degeneracy is clearly broken in the 2C.

We note that the apparent height of 2.1 $\pm$ 0.1~\AA~for the 2F phase, is much smaller than the theoretical value expected for AA stacking of 3.66~\AA.~\cite{cha94} 
However, since STM probes the local density of states (LDOS), the measured apparent height can be significantly different from the interlayer distance if the layers are differently doped. Though the spot size of the photon beam does not allow us measuring the LDOS of the two bilayers separately, ARPES measurements reveal a different charge carrier concentration for the  two bilayers due to a different charge transfer toward the first graphene layer. This different doping between the layers affects the STM measurements and may result in the much lower apparent height of the 2F layer with respect to the 2C layer.
\newline The reasons why this second type of second graphene layer forms, and why it needs higher ethylene exposures to form can be attributed to the following. The difference in energy between the AB and AA staking is only 20~meV/atom,~\cite{lee08, cha94} with AB stacking being more stable. This difference is sufficiently small, such that the AA stacking can be stabilized by kinetic effects which are more dominant under CVD growth with higher supersaturation.

 Finally, we discuss the nature of the replica bands observed for the $\pi_{\rm 2C}$ state. Figure~4(a) show the first derivative of the $\pi_{\rm 2C}$ band dispersion close to the {\rm K} point for the 10$^{2}$~L sample. Replicas may arise by umklapp scattering from the moir{\'e} pattern of the first graphene layer.~\cite{sut09} Note that graphene states also display breaks in the dispersion at the bands crossing (highlighted by red arrows) which reveal energy {\rm minigaps}, similarly to graphene on Ir.~\cite{ple09, rus10, pap12} Indeed, final state scattering enhances the photoemission intensity of the bands and may favor the resolution of the third order replicas by ARPES, however, the umklapp process cannot induce minigaps. These gaps are certainly due to a periodic perturbation to the crystal potential felt by the electrons in the initial state. This is supported by the STM measurement presented in Figure~4(b) revealing that the corrugated layer exhibits a  long-range moir{\'e}  pattern of period $\sim$2.7~nm.~\cite{cal}

{\bf CONCLUSION}

In summary, we show evidence of a new phase of bilayer graphene on the Ru(0001) surface. Following massive ethylene exposure, ARPES spectra reveal a new $\pi$ state
 coexisting with the known ones which is reminiscent of a  different interaction with the first graphene layer. STM investigations show that the Ru surface is fully covered by the first graphene layer which produces a strongly interacting $\pi_{1}$ band. On top of this layer, we find two phases of second graphene layer. One of the two (2C) exhibits Bernal stacking with respect to the first layer. This layer displays weak vertical interaction and a moir{\'e}  pattern with marginal long-range order which modulates the crystal potential. This layer generates a sharp {\it n}-doped Dirac cone ($\pi_{2C}$), replica bands and minigaps. The other phase (2F) exhibits a single Dirac {\it p}-doped ($\pi_{2F}$) state with nearly freestanding character. STM results show that this second layer phase is arranged in an AA stacking sequence with respect to the first layer and has a relatively weak structural and electronic modulation. We thus prove that the first layer can be used as a template to grow different phases of graphene bilayers.
 
{\bf METHODS}

Measurements were performed in three different ultrahigh vacuum (UHV) chambers. 
The presented ARPES measurements were performed at the VUV-Photoemission beamline of the Elettra Synchrotron Facility in Trieste at 120~K, with a photon energy of 100~eV, and an energy resolution of 30~meV. The presented STM experiments were carried out at the Ecole Polytechnique F\'ed\'erale in Lausanne (EPFL) with a home-built low temperature scanning tunneling microscope operating at 0.4~K and used for the present study at 4~K.~\cite{cla11} Additional ARPES and STM measurements were collected at the APE-IOM beamline of Elettra. This beamline has the STM and ARPES spectrometer connected {\it in situ}, so that complementary characterization of the same surfaces can be performed. At this endstation, the primary energy has been set to 65~eV whereas the other parameters have been kept identical, and the home-built STM has been operated at 300~K. All experiments were performed under UHV condition with a base pressure below  $1 \times 10^{-10}$~Torr.
The Ru(0001) crystal was cleaned by repeated cycles of Ar$^{+}$ sputtering ($E = 1200$~eV) and annealing at 1500~K. The order and cleanness of the sample was monitored by low-energy electron diffraction and photoemission spectroscopy at VUV and by STM measurements at EPFL and APE. A graphene bilayer was grown by exposing the Ru(0001) surface held at a temperature of 1600~K to 
$10^{2}$~L (1.9$ \times 10^{-6}$~Torr for 53~s) of ethylene (C$_{2}$H$_{4}$).~\cite{nat12}
The multiphase graphene bilayers were obtained by additional exposure to $10^{5}$~L (1.9$ \times 10^{-4}$~Torr for 530~s) ethylene with the Ru crystal at 1600~K. Samples that have been prepared and characterized at Elettra have been transported to Lausanne and imaged in STM with atomic resolution after a flash to 800~K in a pressure of $1 \times 10^{-10}$~Torr.

{\it Acknowledgment.} This work has been supported by the European Science Foundation (ESF) under the EUROCORES Program EuroGRAPHENE, by the Swiss National Science Foundation, and by the Italian Ministry of Education, University, and Research (FIRB Futuro in Ricerca, PLASMOGRAPH Project).

$^{\ast}${e-mail: marco.papagno@fis.unical.it}

{\bf REFERENCES AND NOTES}

\end{document}